\documentclass[aps,onecolumn,tightenlines,showpacs,aps,amsmath,amssymb,nofootinbib]{revtex4}

\usepackage{latexsym}
\usepackage{graphicx}
\usepackage{subfigure}
\usepackage{cases}
\usepackage{hyperref}
\usepackage{amssymb}
\usepackage{bm}
\usepackage{natbib}
\usepackage{amsmath}

\def\eq{{\rm eq}}
\def\obs{{\rm obs}}

\begin{document}

\title{
21~cm Angular Spectrum of Cosmic String Loops
}

\date{\today}

\author{Hiroyuki Tashiro}
\affiliation{ 
Physics Department, Arizona State University, Tempe, Arizona 85287, USA.
}

\begin{abstract}
The 21~cm signatures induced by moving cosmic string loops are
investigated. Moving cosmic string loops seed filamentary nonlinear
objects. We analytically evaluate the differential 21~cm brightness
temperature from these objects. We show that the brightness temperature
reaches $200$~mK for a loop whose tension is about the current upper limit,
$G\mu \sim 10^{-7}$. We also calculate the angular power spectrum,
assuming scaling in loop distribution. We find that the angular power spectrum for
$G \mu > 10^{-8}$ at $z=30$ or $G\mu >10^{-10}$ at $z=20$ can dominate
the spectrum of the primordial density fluctuations. Finally we show
that a future SKA-like observation has the potential to detect the power
spectrum due to loops with $G \mu =10^{-8}$ at $z=20$.
\end{abstract}


\maketitle

 \section{Introduction}

Cosmic strings are topological defects that could be produced at
phase transitions in the early universe~\cite{Kibble:1976sj} (for
reviews, see Refs.~\cite{VilenkinBook,Hindmarsh:1994re}). Therefore, a
detection or a constraint of cosmic strings can give us direct access to
high energy particle physics and the early universe.

Cosmic strings can produce various observational phenomena, such as CMB
anisotropies~\cite{Kaiser:1984iv,Seljak:1997ii,Allen:1997ag,Albrecht:1997nt,Pogosian:2004ny,Bevis:2007qz,Bevis:2007gh,Pogosian:2007gi,Fraisse:2007nu,Hindmarsh:2009es,Hindmarsh:2009qk,Regan:2009hv,Dvorkin:2011aj,Ringeval:2012tk,Ade:2013xla},
CMB spectral distortions~\cite{Tashiro:2012pp,Chu:2013fqa}, early large
scale structure
formations~\cite{Zeldovich:1980gh,Vilenkin:1981iu,Silk:1984xk,1986MNRAS.222P..27R,Hara:1993ga},
early reionization~\cite{Avelino:2003nn,Pogosian:2004mi,Olum:2006at,
Shlaer:2012rj} and gravitational
waves~\cite{Vachaspati:1984gt,Damour:2000wa,Damour:2001bk,Damour:2004kw,Olmez:2010bi,vanHaasteren:2011ni,Sanidas:2012ee,Binetruy:2012ze}.
Since most observational effects due to strings are gravitational, we
can set observational constraints on the strength of gravitational
interactions of strings which is parametrized by the dimensionless
number $G \mu$, where $G$ is Newton's constant and $\mu$ is the mass per
unit length (or tension) of string.  The current limit on $G \mu$ is
obtained from CMB anisotropy observations.  WMAP and SPT data provide
the limit $G\mu <1.7 \times 10^{-7}$~\cite{Dvorkin:2011aj}.  Recently
Planck data updated the limit to $G\mu <1.5 \times
10^{-7}$~\cite{Ade:2013xla}.

Cosmological 21~cm observation is expected as  
one of the new observational windows for cosmic strings.
Since the intensity of the redshifted 21~cm lines of neutral hydrogen
is sensitive to the number density and
temperature of neutral hydrogen,
cosmological 21~cm observation has the potential to probe the high redshift
Universe through the epoch of reionization ($8<z<20$) to the dark ages ($z > 20$)
(for reviews, see Refs.~\cite{Furlanetto:2006jb,Pritchard:2011xb}).
Additionally, choosing the observational frequency, we can obtain the
three-dimensional map of the redshifted 21~cm intensity \cite{1997ApJ...475..429M}.
Therefore, cosmological 21~cm observation is suitable for searching the signatures
of the structure formations due to cosmic strings.
Currently, there are some ongoing and future projects for measuring the
cosmological 21~cm radiation;
MWA\footnote{http://www.haystack.mit.edu/ast/arrays/mwa/},
LOFAR\footnote{http://www.lofar.org/},
GMRT\footnote{http://gmrt.ncra.tifr.res.in},
PAPER\footnote{http://astro.berkeley.edu/˜dbacker/},
SKA\footnote{http://www.skatelescope.org/}
and Omniscope~\cite{Tegmark:2009kv}.

Cosmological 21~cm signatures due to cosmic strings have been
investigated in several papers~\cite{Khatri:2008zw,
Berndsen:2010xc,Brandenberger:2010hn,
Hernandez:2011ym,Hernandez:2012qs}.  It is well-known that
a cosmic string produces overdense regions by inducing 
gravitational collapse of baryons onto the string wake. Moreover, if
the tension of the cosmic string exceeds a critical value, the collapsed
gas is heated by the collapse shock. As a result, the cosmic wake can
produce strong signature in redshifted 21~cm maps, even if the tension of the
string is smaller than the current
constraint, $G\mu \sim 10^{-7}$~\cite{Brandenberger:2010hn, Hernandez:2012qs}.  The angular
21~cm power spectrum due to string wakes has also been investigated in
Ref.~\cite{Hernandez:2011ym}.

In this paper, we study cosmological redshifted 21~cm signature and its
angular power spectrum due to cosmic string loops.  Loops are produced
through self-interactions and intercommunications of long
strings. Since loops can create gravitational fields, they serve as
seeds for nonlinear structures.  The 21~cm signature due to a static
loop have been studied in Ref.~\cite{Pagano:2012cx}.  A loop induces a
spherical gravitational collapse of matter and produces a bright spot
on a redshifted 21~cm map. The brightness temperature of the spot
reaches 1 K for $G \mu \sim 10^{-7}$.  However loops are expected to have relativistic
initial velocity when they are produced~\cite{Vachaspati:1984dz,
Allen:1990tv}. The non-zero initial velocity makes the collapsed object have
a filamentary structure. Compared with the spherical collapse case due to a
static loop,
the density contrast inside the object is small, and
the resultant shock heating at the collapse is
not efficient. However the
signature is elongated on a map.
We compute the gas temperature in the filament due to a moving loop and evaluate the 21~cm signature.
We also calculate the angular power spectrum, assuming scaling
in loop distribution~\cite{Vanchurin06, Ringeval07,Olum07}.
  
This paper is organized as follows. In Sec.~II, we briefly review the
the accretion onto a moving loop.
In Sec.~III, we investigate the differential brightness temperature of
redshifted 21~cm lines
induced by a loop.
In Sec.~IV, assuming the loop number density distribution, we evaluate the angular 21~cm power spectrum. We show the
angular power spectrum for different values of $G\mu$ and redshifts.
We conclude in Sec.~V.
Throughout this paper, we use natural units ($\hbar = c=1$), and assume
$\Lambda$CDM cosmology with $\Omega_M=0.26$, $\Omega_B=0.05$
and $h=0.7$, which are consistent with the WMAP 9-year results~\cite{Hinshaw:2012aka}.

 \section{accretion onto a moving loop}

A loop can become a seed of structure formations.
Matter accretes
onto a loop, following the gravitational field due to the loop.  In this section, we briefly review the
evolution of the structure produced by a loop (for details, see
Ref.~\cite{VilenkinBook}).

First we assume that a loop is formed at time $t_i$ with length
$L=\alpha t_i$ and an initial velocity $v_i$.
Since loops are non-relativistic objects, their velocity evolution is given by
\begin{equation}
 v(t, t_i) = v_i { a_i \over a(t)},
\end{equation}
where $a(t)$ is the scale factor and the subscript $i$ denotes the value at $t_i$.
Throughout this paper,  we take the values, $\alpha =0.1$ and $v_i = 0.3$, suggested by simulations~\cite{BlancoPillado:2011dq}.
The matter accretion occurs in the matter dominated era, while, in the radiation dominated era, the
accretion is prevented because of the radiation
pressure.
Therefore,
the mater accretion starts at time
\begin{equation}
 t_s (t_i) =
\left\{
\begin{array}{cc}
t_{\eq}, \quad{\rm  for} \quad t_i< t_{\eq} \\
 t_{i},{~} \quad{\rm  for} \quad t_i > t_{\eq}
\end{array}
\right. ,
\end{equation}
where $t_{\eq}$ is the time of matter-radiation equality.

\begin{figure}
\begin{center}
 \includegraphics[width=90mm]{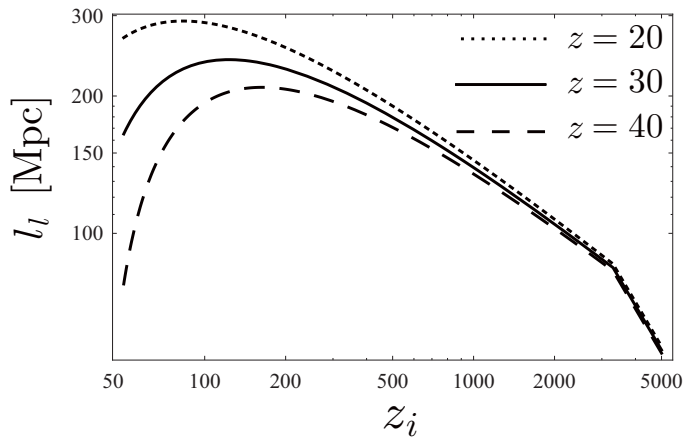}
  \end{center}
  \caption{The comoving length of the filament $l_l$
 as a function of the redshift $z_i$ corresponding to the time $t_i$
 at which loops with $L=\alpha t_i$ are formed.
 The dotted, solid and dashed lines represent $l_l$ at $z=20$,
 $z=30$ and $z=40$, respectively.}
  \label{fig:comoving_l}
\end{figure}

When the loop length $L$ is smaller than the horizon
scale, the gravitational field due to the loop at time $t$ is described as the field
due to the point mass with mass $\mu L$ and velocity $v(t, t_i)$.
Accordingly the accretion is axially symmetric along the direction of the velocity, and the
resultant overdense structure is like a filament.
The comoving length of the filament structure, $l_l$, corresponds to the
comoving length of the loop trajectory after $t_s$,
\begin{equation}
 l_l(t, t_i) =  3 { v_s  t_s \over a_s}
 \left(1
-\sqrt{a_s \over a(t)}
\right),
\end{equation}
where the subscript $s$ denotes the value at $t_s$.
In Fig.~\ref{fig:comoving_l},
we plot the dependence of $l_l$ on the initial
redshift $z_i = z(t_i)$ at which a loop with length $\alpha t_i$ is
produced.
As the redshift $z_i$ increases, the length of the filament $l_l$ decreases.
In particular, the length $l_l$ is strongly suppressed for loops which
are produced at $z_i >z_{\rm eq}$, because such loops do not have
large velocities at time $t_s$, where matter accretion starts.

\begin{figure}
\begin{center}
 \includegraphics[width=90mm]{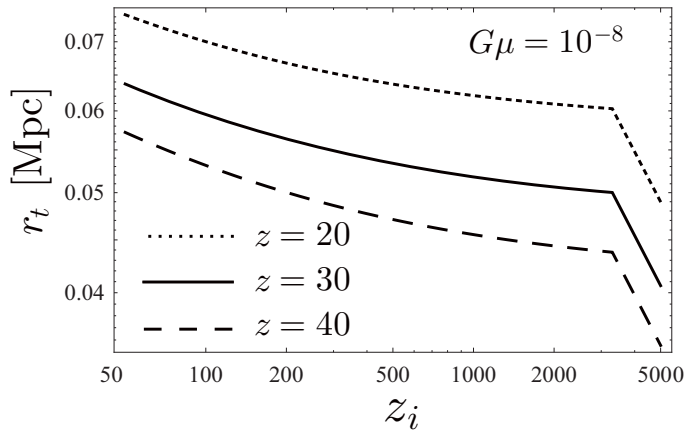}
  \end{center}
  \caption{The comoving turnaround surface of the filament $r_{\rm t}$ as a
 function of the redshift $z_i$.
In this figure, we set $G \mu=10^{-8}$ and $l = l_l/2$.
 The dotted, solid and dashed lines represent $r_{\rm t}$ at $z=20$,
 $z=30$ and $z=40$, respectively. The turnaround surface $r_{\rm t}$ is
 proportional to $\sqrt {G\mu}$.}
  \label{fig:comoving_r}
\end{figure}

The accretion evolution on a loop with the initial velocity has
been studied by using the Zel'dovich approximation in the cylindrical
coordinates $(r, \phi, l)$, where the $l$--axis corresponds to the
direction of the velocity.
The turnaround surface $r_{\rm t}$ on the $r$--axis is obtained by
solving the equation where $r_{\rm t}$ appears on both sides~\cite{1987ApJ...316..489B}, 
\begin{equation}
 r_{\rm t} = 2 d(t_i) f(t, t_i) g_r (r_{\rm t}, t_i,l).
  \label{eq:turnaround}
\end{equation}
Here
\begin{equation}
d(t_i)= { 3 v_s t_{s} },
\quad
f(t,t_i) = {1 \over 5} {G \mu L \over v_s^2 d(t_i)} {a(t) \over a_s} ,
\quad
g_r (r,  t_i, l)= {R_f - R_i \over r} +{l d(t_i) \over r  R_i},
\end{equation}
where $R_i = (r^2 +l^2)^{1/2}$ and $R_f =[r^2+ (l-d(t_i))^2]^{1/2}$.
For our region of interest of $(t, t_i)$, $f(t, t_i)$ is much less than one. In
this limit, the solution of Eq.~(\ref{eq:turnaround}) is given by an
approximated analytic form,
\begin{equation}
 r_{\rm t }(t, t_i,l) =\sqrt{ 4 f(t,t_i) d(t_i) (d(t_i) -l)} /a_s.
 \label{eq:ta_approx}
\end{equation}
Fig.~\ref{fig:comoving_r} shows the dependence of $r_{\rm t}$ on
$z_i$. As the universe evolves, the turnaround surface becomes large.
In Eq.~(\ref{eq:ta_approx}),
$G \mu$ dependence appears only in $f(t, t_i)$.
Therefore $r_{\rm t}$ is proportional to $\sqrt {G \mu}$.
The resultant filament structure is very elongated because
\begin{equation}
 { l_l (t, t_i) \over r_{\rm t}(t, t_i) } \sim 1000 \left(
  { G \mu \over 10^{-8} } \right)^{-1/2} \quad
  {\rm at \ } t(z=30).\label{eq:ratio_xy}
 \end{equation}

\begin{figure}
\begin{center}
 \includegraphics[width=90mm]{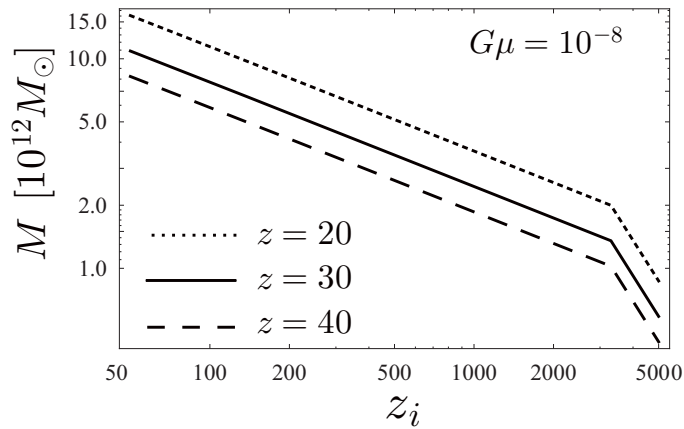}
  \end{center}
  \caption{The filament mass dependence on $z_i$.
 In this figure, we set $G \mu=10^{-8}$.
 The dotted, solid and dashed lines represent the filament mass at $z=20$,
 $z=30$ and $z=40$, respectively. The mass of the filament is
 proportional to $G\mu$.}
  \label{fig:m_solar}
\end{figure}

The accreted mass at time $t$ corresponds to the mass inside $r_{\rm t}(t) $ along
the trajectory $l_l(t)$ in the comoving frame,
 \begin{equation}
  M(t, t_i) \approx  \rho_M \int _0 ^d \pi r_{\rm t} (t) ^2 dl =
{3 \over 5}  \mu L {a(t) \over a_s},
\label{eq:mass_filament}
 \end{equation}
 where $\rho_M $ is the comoving matter density. 
Eq.~(\ref{eq:mass_filament}) tells us that the accreted mass is proportional to $G \mu$.
We plot the accreted mass in Fig.~\ref{fig:m_solar}.
The accreted mass grows as the universe evolves.
Since the mass depends on the loop length $L$, the mass is proportional
to $z_i ^{3/2}$. However, since the loop generated before $z_{\rm eq}$ cannot
produce the filament until $z_{\rm eq}$, the resultant filament mass is
suppressed for $z>z_{\rm eq}$. 

\begin{figure}
\begin{center}
 \includegraphics[width=90mm]{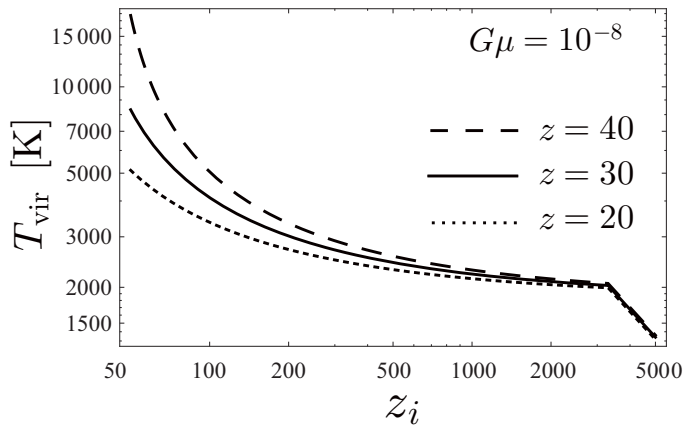}
  \end{center}
  \caption{The virial temperature of the filament as a function of the
 redshift $z_i$
In this figure, we set $G \mu=10^{-8}$.
 The dashed, solid and dotted lines represent $T_{\rm vir}$ at $z=40$,
 $z=30$ and $z=20$, respectively. The virial temperature is
 proportional to $G\mu$.}
  \label{fig:Tvir}
\end{figure}

The gas inside the turnaround surface $r_{\rm t}$ collapses and is virialized.
The collapsed object shrinks to about half of the physical
turnaround surface scale.
Although dark matter condenses in the center of the object,
baryon suffers the virialization shock due to the collapse.
As a result, the baryon distribution is almost homogeneous and isothermal
at the virial temperature.
The virial temperature of the filament structure is given by
\cite{Eisenstein:1996pr} 
\begin{equation}
 T_{\rm vir} \approx {1 \over 2} m_p G \lambda_{\rm fil}.\label{eq:virtemp}
\end{equation}
Here $m_p$ is the proton mass and $\lambda_{\rm fil }$ is a linear mass
density of the filament structure,
\begin{equation}
 \lambda_{\rm fil} \sim {M(t,t_i) \over a(t) l_l (t,t_i)},
\end{equation}
where $a (t ) l_l(t, t_i) $ represents the physical length scale of the filament.
Since $M$ is proportional to $G\mu$ and $l_l$ does not depend on $G\mu$,
$T_{\rm vir}$ is proportional to $G \mu$.
We plot the evolution of the virial temperature of the filament in Fig.~\ref{fig:Tvir}.

We assume that the virialized radius is half of the physical turnaround
surface scale. Accordingly, the matter density contrast inside the filament is 
\begin{equation}
 \rho_f = {4  \rho_{M} } ,\label{eq:deltadensity}
\end{equation}
where $\rho_M$ is the background matter density.
Although the turnaround surface depends on $l$ as shown
Eq.~(\ref{eq:ta_approx}), for simplicity,
we introduce 
the typical turnaround surface $\bar r_{\rm t}$ which corresponds the turnaround surface at $l=l_l/2$.

\section{21~cm signature from the accreted filament}

As shown in the previous section,
the gas density and temperature inside a filament are different from
those of background values. This difference makes the optical depth of
the 21 cm transition in the filament different from one in the
intergalactic medium (IGM). As a result, we can
observe the filament due to the loop in the differential brightness
temperature map of the redshifted 21~cm line.

The optical depth of a filament to the photon at the frequency $\nu$
 along the Line Of Sight~(LOS) is calculated as the one of a virialized object~\cite{Furlanetto:2002ng},
\begin{equation}
\tau(\nu) = {3  A_{10}   \over 32 \pi k \nu_*^2 } \int d R ~
  {x_{\rm HI}(R) n_{H } (R ) \over T_s(R) } \phi(\nu),
  \label{eq:def_opt}
\end{equation}
where $\nu_*$ is the hyperfine transition frequency, $\nu_*=1420.4~$MHz,
$A_{10} = 2.85 \times 10^{-15} {\rm s}^{-1}$ is the spontaneous
emission rate, $x_{\rm HI}$ is the neutral fraction of hydrogen, $T_s$ is
the spin parameter, and $\phi(\nu)$ is the intrinsic line profile.

Now we focus on the filament whose virial temperature is smaller than
$10^4~$K. When the virial temperature exceeds $10^4~$K,
the atomic cooling is efficient enough to cause the further collapse and
the star formations occur in the filament structure. Once stars form, they ionize the
surrounding hydrogen by emitting UV photons. However, below $10^4~$K,
no sources of UV photons are produced inside. Therefore, we set the neutral fraction 
$x_{\rm HI} =1$ inside the filament.

The spin temperature is related to the ratio between the neutral
hydrogen number density in the excited and ground states of the hyperfine structure.
Without stars, the excitation or de-excitation of the hyperfine
structure is caused by the thermal kinetic collision at
the virial temperature, the spontaneous
emission and the stimulated emission by CMB photons.
Therefore the spin temperature of the filament is obtained as
\cite{1958PIRE...46..240F, 1959ApJ...129..536F} 
\begin{equation}
 T_s = {T_\gamma +y_k T_{\rm vir} \over 1+y_k}.
\end{equation}
Here $y_k$ is the kinematic coupling term \cite{Kuhlen:2005cm},
\begin{equation}
 y _k = {T_* n_{\rm Hf} \kappa \over A_{10} T_{\rm vir}},
\end{equation}
where $T_* = 0.068~$K, $\kappa$ is the effective single-atom rate coefficient;
$\kappa = 3.1 \times 10^{-11} T^{0.357} \exp (-32/T ) {\rm cm}^3 {\rm
s}^{-1}$, and $n_{\rm Hf}$ is the neutral hydrogen number density of the filament, 
$n_{\rm Hf} = \rho_f \Omega_B/ (m_p \Omega_M) $.

\begin{figure}
\begin{center}
 \includegraphics[width=90mm]{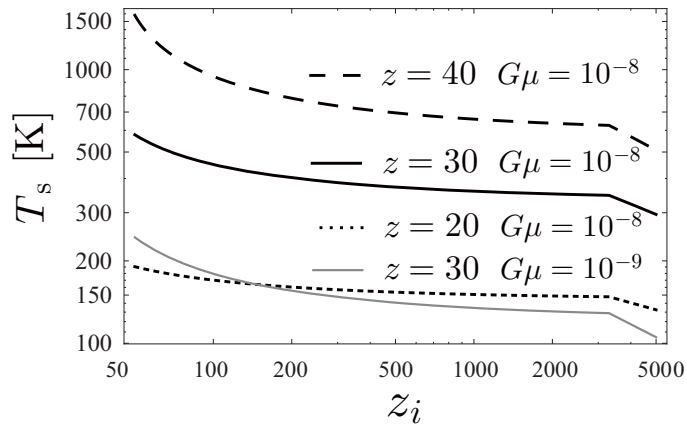}
  \end{center}
  \caption{The spin temperature of the filament
as a function of $z_i$.
 The dashed, solid and dotted lines represent $T_s$ at $z=40$,
 $z=30$ and $z=20$ with $G\mu =10^{-8}$, respectively.
 We also plot $T_s$ at $z=30$ with $G \mu=10^{-9}$
as the gray line.}
  \label{fig:Tspin}
\end{figure}

The integration in Eq.~(\ref{eq:def_opt}) is performed along the LOS.
Since our filament model is isothermal and has the uniform density profile,
the integration can be replaced by the column density of the filament
along the LOS.
As discussed in the previous section,
the loop with length $L$ produces the filament whose comoving typical
radius
is $(1/2) \bar{r}_{\rm t}$. 
We define the impact parameter $b$ from the symmetrical axis in the unit of $(1/2) \bar{r}_{\rm t}$. The width of the filament along the LOS
depends on the impact parameter and the angle between the LOS direction and
the symmetrical axis of the filament, $\psi$.
Therefore, the column hydrogen number density of the
filament at redshift $z$ with the
impact parameter $b$ is written as
\begin{equation}
 N_{\rm H} (b)= n_{H f } {a \bar{r} _{\rm t} \over \sin \psi} \sqrt{1- {b^2 \over
 a ^2 }},
 \label{eq:column_density}
\end{equation}
where $a$ is the scale factor at $z$.
Note that, since the filament has the finite length, the column number
density cannot exceed $a n_{H f } l_l$. Therefore we set this upper limit
on the column number density.

With the column hydrogen number density, the optical depth of the
 filament to the photon at the frequency $\nu$ along
 the LOS with the impact parameter $b =0$ and $\sin \psi=1$ is given by
\begin{equation}
\tau_0 (\nu) = {3  A_{10}   \over 32 \pi  \nu_*^2 }
  { N_{\rm H } (0) \over k_B T_s } \phi(\nu).
\end{equation}

The intrinsic line
profile is broadened by the Doppler effect, because the gas inside the filament has high temperature. We adopt the Doppler
broadened form,
\begin{equation}
\phi _D(\nu ) = (\Delta \nu \sqrt \pi)^{-1} \exp \left [ - (\nu -\nu_*)^2
/\Delta \nu^2\right ] ,
\end{equation}
with $\Delta \nu = \nu_* \sqrt{2 k_B T_{\rm vir}
/ m_{\rm H} }$.
The Hubble flow also causes the line broadening as considered in Ref.~\cite{Pagano:2012cx}.
The redshift difference along the LOS is given
as $\delta \nu = 2 H a x_t / \sin \psi$.
As a result, the intrinsic line
profile broadened by the Hubble flow is~\cite{Pagano:2012cx}
\begin{equation}
 \phi _H(\nu ) =
  \left\{
  \begin{array}{ll}
1/\delta \nu \quad&  |\nu -\nu_*| < \delta \nu /2 \\
0 \quad & {\rm otherwise} \\
  \end{array}
\right. .
\end{equation}
We take into account only the largest effect between them.
In most of our cases, the Doppler effect dominates the Hubble flow effect.

\begin{figure}
\begin{center}
 \includegraphics[width=90mm]{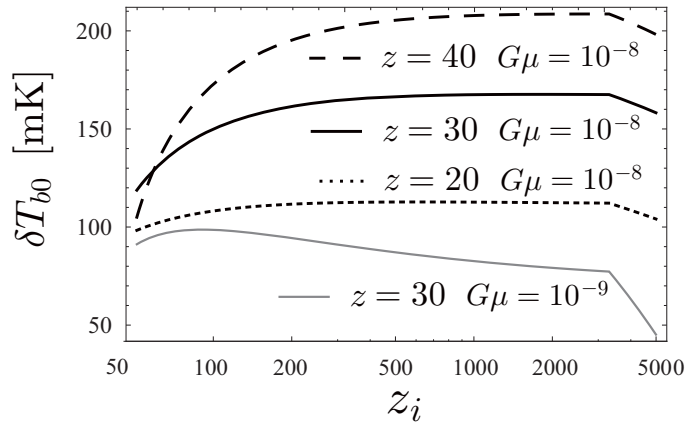}
  \end{center}
  \caption{The brightness temperature of the filament $\delta T_{b0}$
 as a function of $z_i$. 
 The dashed, solid and dotted lines represent $\delta T_{b0}$ at $z=40$,
 $z=30$ and $z=20$ with $G\mu =10^{-8}$, respectively.
 We also plot $\delta T_{b0}$ at $z=30$ with $G \mu=10^{-9}$
as the gray line.}
  \label{fig:dTemp}
\end{figure}

The observable quantity is the differential brightness temperature of the 21~cm signal.
In the case with $\tau \ll 1$, 
the differential brightness temperature of the filament with $b=0$ and $\sin \psi=1$
at $z$ is given by 
\begin{equation}
 \delta T_{b0} (\nu_{\rm obs}) = {T_s -T_{\gamma}(z)  \over 1+z}
  \tau_0(\nu_{\rm obs}(1+z)),
\label{eq:brightness_21_0}
\end{equation}
where $\nu_{\rm obs}$ is the observation frequency; $\nu_{\rm obs} = (1+z) \nu_*$. We plot the brightness
temperature $\delta T_{b0}$ in Fig.~\ref{fig:dTemp}. 

The loops produced from $z \sim 100$ to $1000$ can create the signal
$\delta T_{b0} \sim 150$~mK around $z=30$.
As $G\mu$ increases, the signal becomes stronger.
Although the filaments due to large loops have larger spin
temperature, their differential brightness temperature $\delta T_{b0}$ is
strongly damped. Since such filaments have large virial temperature, the
Doppler broadening is effective enough to suppress $\delta T_{b0}$.

It is worth pointing out that
our values are smaller than the
ones in Ref.~\cite{Pagano:2012cx} in which a loop produces the
brightness temperature as high as 1~K for $G \mu \sim 10^{-7}$.
This is because they have considered static loops. Static
loops produce spherical collapsed objects. The density contrast in
such objects is roughly 64, while the one in the filament object is
4. Therefore, a static loop can produce larger signal than a moving
loop.

According to Eq.~(\ref{eq:brightness_21_0}),
the sign of $\delta T_{b0} $ depends on the difference between $T_s$ and
$T_\gamma$. As shown in Fig.~\ref{fig:Tspin}, most of the filaments due to
loops with $G \mu =10^{-8}$ have spin temperature larger than the CMB
temperature.
Therefore, the 21~cm signals from such filaments are observed as 
emissions for the CMB, namely $\delta T_{b0} >0$. 
On the other hand, the loops with $G \mu =10^{-9}$ produce filaments
whose spin temperature is lower than $T_s$ for $G \mu = 10^{-8}$. In
particular, the spin temperature of the filaments produced by small
loops generated before $z_{\rm eq}$ is strongly suppressed. As a result, the
spin temperature becomes smaller than the CMB temperature, and the
signal of the filaments which have such lower temperatures are observed
as absorptions for the CMB.

Following Eq.~(\ref{eq:column_density}), we can extend
Eq.~(\ref{eq:brightness_21_0}) to 
the differential brightness
temperature with the impact parameter $b$ and the
angle $\psi$,
\begin{equation}
 \delta T_{b} (\nu_{\rm obs} , b , \psi) =  
{ \delta T_{b0} (\nu_{\rm obs} )   \over \sin \psi} \sqrt{1- {b^2 \over
  a^2 }}.
\label{eq:brightness_21}
\end{equation}

Since filaments due to loops have non-zero differential brightness
temperature, the filaments are observable by 21~cm observations.
We can observe the filament as a projection on the 2-D angular map of the celestial sphere. 
The projected filament with the angle $\psi$ of the LOS at the
redshift $z$ is a rectangle with the angular scales of the sides,
$\theta_x =r_t  / (2 D_A)$ and $\theta_y =l_l \sin \psi /D_A$
 where $D_A$ is the comoving angular diameter distance to $z$.
For the redshift $z=30$, $\theta_x$ corresponds to $\sim 1''$. The
angular resolution of the current design of the planned cosmological
21~cm observations is much larger than this angle scale.
Therefore, it is difficult to resolve an individual filament due to a loop by the
21~cm observations.

However filaments can contribute to the angular power spectrum on the
observable scales of the 21~cm observations
as the tail of the angular power spectrum of the filaments on large
scales. In the next section, we evaluate the angular power spectrum
of the 21~cm signals from the filaments due to loops. 

\section{angular power spectrum}

In order to calculate the angular power spectrum of 21~cm signals from
the filaments due to loops, we follow the analytical approach to
calculate the the Sunyaev--Zel'dovich effect due to large-scale filament
structures~\cite{shimon:2012vs}.
The total angular power spectrum due to nonlinear structures
can be separated into two parts \cite{1988MNRAS.233..637C}.
The first component is a Poisson term which is the contribution
from the correlation between two points in the same structure.
The second is a clustering term
due to the correlations between different structures. 
We assume that the spatial distribution of loops is totally random, and
there is no correlations between different filaments.
Therefore,
the angular power spectrum due to loops can be expressed
only by the Poisson term,
\begin{equation}
 C_\ell = \int dz 
{d V (z)\over dz} 
 \int dL \int d\psi {dn(L, z) \over dL} f(\psi)  W(\nu_\obs, z)^2  P ( {\ell}, L, \psi),
\label{eq:power_cl}
\end{equation}
where $V(z)$ is a comoving volume
element per steradian at a redshift $z$, ${dn (L, z) / d L}$ is
the comoving number density per unit length of loops of length $L$ at $z$,
$f(\psi)$ is the probability function for the angle between the LOS
and the symmetrical axis of the filament, $\psi$, and $P ( {\ell}, L, \psi)$
is the 2-dimensional power spectrum of the 21~cm signal from a
single filament produced by a loop of length $L$ and angle
$\psi$, which we shall discuss later.
In Eq.~(\ref{eq:power_cl}), 
$W(\nu_\obs,z)$ is a response function associated with
bandwidth of the experiment.
Generally, $W(\nu_\obs,z)$
is a function of the frequency centered at the observed
frequency $\nu_\obs$. Since 
there is one to one correspondence between the frequency
and the redshift, the response function $W(\nu_\obs,z)$ has a peak
at the redshift $z$ which satisfies $\nu_\obs = \nu_*/(1+z) $. For simplicity, we take
$W(\nu_\obs,z)$ to be flat with the width $\Delta z=1$ in redshift space.
We set $f(\psi) = (1/2) \sin \psi $, which is based on the assumption that the probability
is proportional to the solid angle element.

The comoving number density of cosmic
string loops of length $L$ in comoving volume at the matter dominated
epoch is
\cite{VilenkinBook} 
\begin{equation}
{{dn(L,z) \over dL }=\frac{\kappa_{L}}{(1+z)^3 p} } 
\frac{ C_L }{t(z) ^2 L^2},
\label{dnloop}
\end{equation}
where $\kappa_{L} \sim 2$~\cite{BlancoPillado:2011dq}, $p\leqslant 1$ is the
reconnection probability, $C_L$ is $\sqrt{{t_{\rm eq}}/L}$ for $L< \alpha t_{\rm eq}$ and $1$ for $L>\alpha t_{\rm eq}$.
In Eq.~(\ref{dnloop}) we ignore the decay of the loops 
due to gravitational radiation
because this is not important for the loops which 
we consider in this paper.

\begin{figure}
  \begin{center}
   \includegraphics[width=70mm]{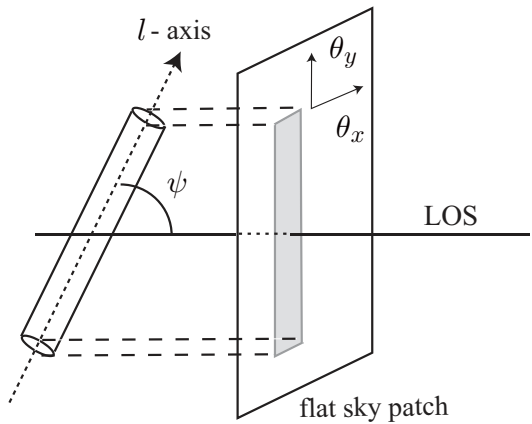}
  \end{center}
\caption{The projected filament on the flat sky patch ($\theta_x, \theta_y$) is represented in gray.
The observer is at the right end on the LOS.
The filament has an inclination angle $\psi$ between the symmetrical
 axis (the $l$--axis) and
 the LOS.}
\label{fig:frame}
\end{figure}

The 2-D spectrum of a single filament can be calculated from Eq.~(\ref{eq:brightness_21}).  
Since the angular size of the projected filament on a 21~cm map is
 small, we consider the projected filament on a small flat sky
 patch $(\theta_x , \theta_y)$ where the $\theta_x$ (or $\theta_y$) axis is
 normal to (or along)
 the symmetrical axis of the projected filament as shown in Fig.~\ref{fig:frame}.
The angular power spectrum of the filament
 corresponds to the 2-D Fourier power spectrum in the flat space in this approximation.
 The profile of the 21 cm brightness temperature of the filament on the
 small flat sky patch is
 given by
\begin{equation}
 \delta T_{b} ( \theta_x, \theta_y) =  
{ \delta T_{b0}   \over \sin \psi}
\sqrt{1- {4 D_A^2
\theta_x ^2 \over  r_{\rm t}^2 }} 
~\Theta \left({r_{\rm t} \over 2 D_A }  - |\theta _x| \right)
~\Theta \left({l_l \sin \psi \over 2 D_A 
}  - |\theta _y| \right),
\label{eq:brightness_21_small}
\end{equation}
where $\Theta (x)$ is the unit step function.
Depending on the filament length and the angle $\psi$ between the LOS
and the symmetrical axis of the filament, 
the distance between the head and tail of the filament in the redshift
direction may be larger than the width of the response function.
In this case, the whole filament cannot be observed in the same redshift
bin, and we must take into account this effect in Eq.~(\ref{eq:brightness_21_small}).
However, we confirmed that most of the contributions to the angular
power spectrum are due to small filaments whose sizes are smaller than
the width $\Delta z = 1$. Therefore, Eq.~(\ref{eq:brightness_21_small}) is
valid in our calculation with $\Delta z =1$.

We perform the Fourier transform of Eq.~(\ref{eq:brightness_21_small}) in the 2-D flat space and we obtain
the Fourier component,
  \begin{equation}
 \delta  \tilde T_{b} ( \ell_x, \ell_y) =  
{2 \pi  \delta T_{b0}    \over \ell_x \ell_y \sin \psi}
{J_1 \left({\ell_x r_{\rm t} \over 2 D_A }\right)  }
 \sin \left( {\ell_y l_l \sin \psi  \over 2 D_A} \right) .
  \end{equation}

Accordingly, the 2-D spectrum is simply given by
\begin{equation}
 P_2 (\ell_x, \ell_y)
  =\left[
  {2 \pi  \delta T_{b0}    \over \ell_x \ell_y \sin \psi}
{J_1 \left({\ell_x r_t \over 2 D_A }\right)  }
 \sin \left( {\ell_y l_l \sin \psi  \over 2 D_A} \right) \right]^2.
\end{equation}  

In order to get the power spectrum $P(\ell, L, \psi)$ at a given $\ell$, we must take into
account the contributions from 
all $(\ell_x,\ell_y)$ which satisfy $\ell^2 = \ell^2 _x +\ell^2 _y $,
\begin{equation}
 P(\ell, L, \psi ) = {1 \over \ell} \sum_{(\ell _x , \ell
 _y)} P_2 (\ell_x, \ell_y).
\end{equation}

\begin{figure}
 \begin{tabular}{cc}
 \begin{minipage}{0.5\hsize}
  \begin{center}
   \includegraphics[width=80mm]{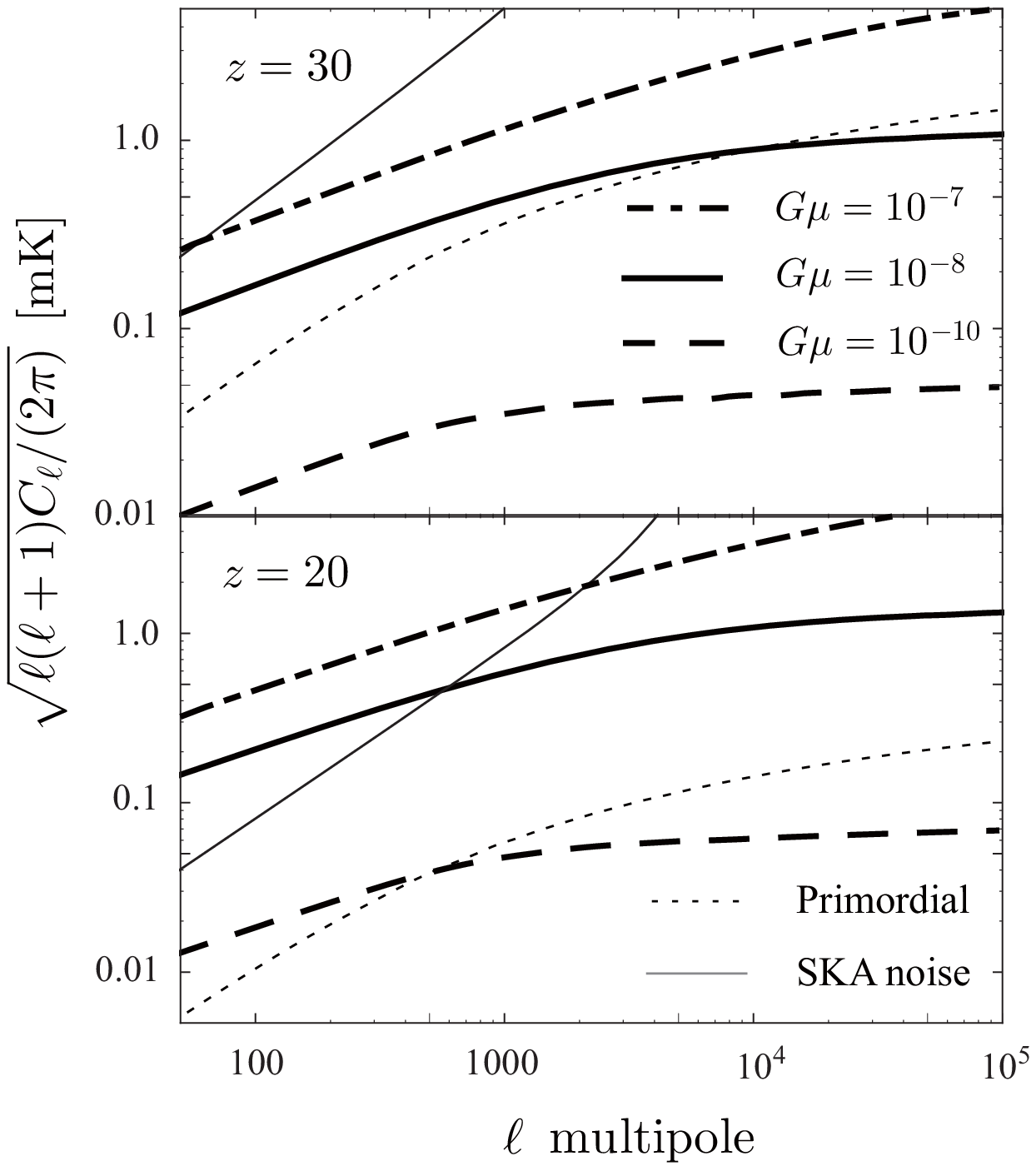}
  \end{center}
 \end{minipage}
 \begin{minipage}{0.5\hsize}
  \begin{center}
  \includegraphics[width=80mm]{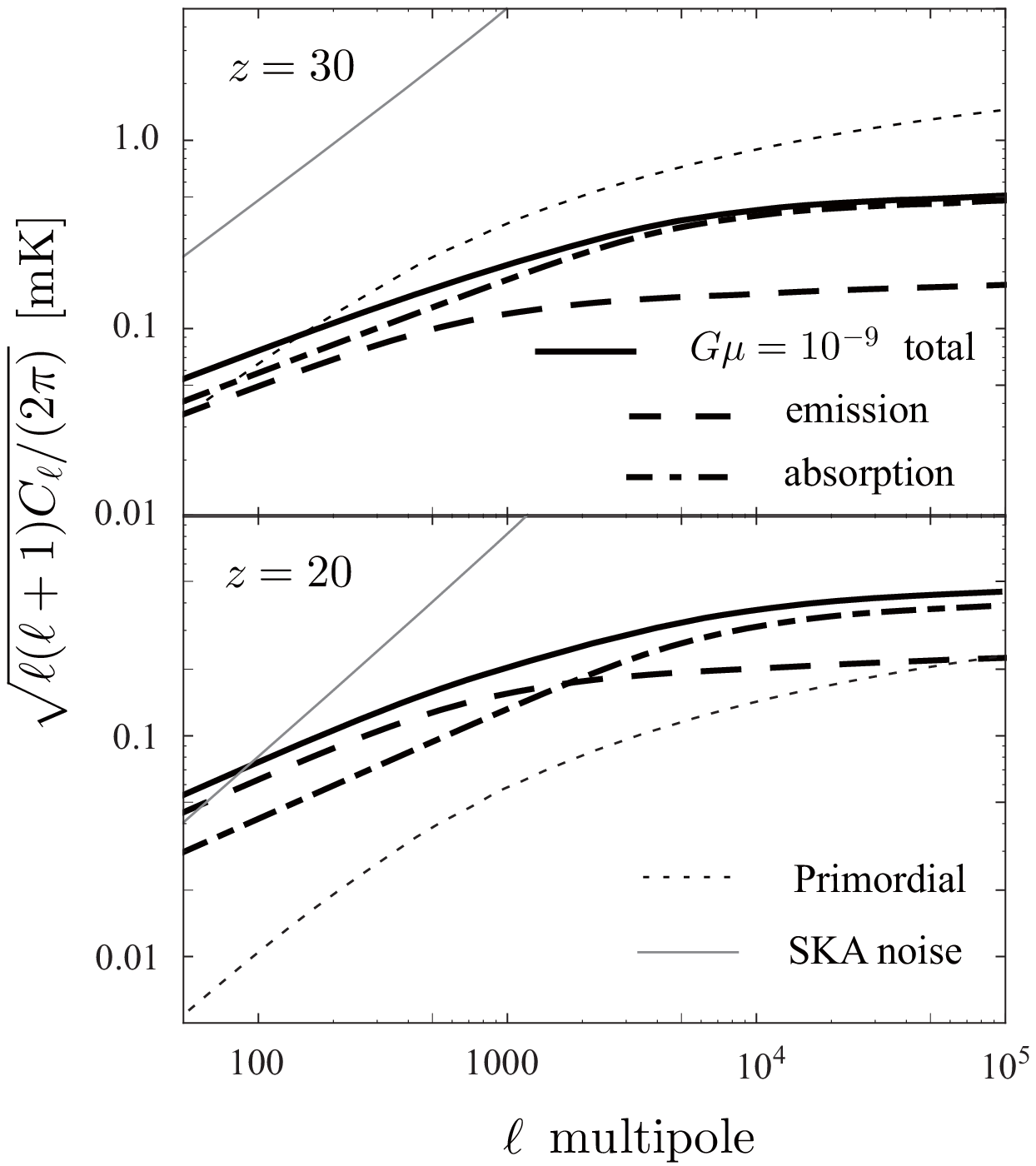}
  \end{center}
 \end{minipage}
\end{tabular}
  \caption{The angular power spectrum due to loops.
In the left panel, we plot the
 spectra with $G\mu=10^{-7}$, $G\mu =10^{-8}$ and $G\mu = 10^{-10}$ as
the dot-dashed, solid and dashed lines, respectively. 
In the right panel, 
the solid line represents the total angular power spectrum for $G \mu
 =10^{-9}$, and
the dashed and the dot-dashed lines show contributions from emission
 and absorption components, respectively. 
 From top to
 bottom in both panels, we set $\nu_\obs$ corresponding to $z=30$,
 and $z=20$.
For comparison, we show the spectrum due to the primordial density
 fluctuations with the dotted line. We
 also plot the noise power spectrum of the SKA-like observation as the
 gray line.}
  \label{fig:llcl_z}
\end{figure}

Using Eq.~(\ref{eq:power_cl}),
we calculate the angular power spectrum for different $G \mu$.
We represent the results of 
the angular power spectra in Fig.~\ref{fig:llcl_z}.

The amplitude of the spectrum depends on $G\mu$. Larger $G \mu$ produces
higher amplitude of the angular power spectrum. 
The angular power spectrum due to loops does not have a strong redshift
dependence.  At small $\ell$, the spectrum $\ell^2 C_\ell$ is
proportional to $\ell$.  However, the slope becomes shallow with
increasing $\ell$, and the spectrum is almost constant at large $\ell$.  The
multipole at which the slope changes corresponds to the typical length
scale of the filaments.  For larger $G \mu$, since even small filaments
can contribute to the spectrum, the multipole at which the slope changes
shifts to larger $\ell$.

We found that most contributions come from filaments produced by small
strings generated $z> z_\eq$. The length of such filaments is smaller than
the width of the response function, $\Delta z =1$. Hence, as mentioned above, 
Eq.~(\ref{eq:brightness_21_small}) is valid in our calculation.

The promising signal of the cosmological 21~cm fluctuations is provided by the
primordial density fluctuations. For reference, we plot the
angular power spectrum due to the primordial density fluctuations obtained through CAMB
for the sharp window function~\cite{Lewis:2007kz}.
At $z>30$, the angular power spectrum for $G \mu > 10^{-8}$
can dominate the primordial fluctuations.
Although the angular power spectrum induced by loops is almost independent of the
redshift, the amplitude due to the primordial fluctuations becomes small as
the redshift decreases.
As a result, the spectrum due to loops
dominate the primordial density fluctuations
at low multipoles $\ell < 1000$ at $z=20$ even for $G \mu = 10^{-10}$.

In the power spectrum for $G\mu > 10^{-8}$,
most of the contributions are emission signals. On the other hand, the absorption signals
dominate for $G \mu = 10^{-10}$. This is because small $G \mu$ cannot
produce massive filaments in which the filament gas heats up to the
virial temperature larger than the CMB temperature.
For $G \mu = 10^{-9}$, absorption and emission contributions are comparable.
Since the filament virial temperature grows with time, the absorption contribution becomes
large at low redshifts.
The 21~cm signals due to the primordial density fluctuations
before the epoch of reionization are absorption signals. Hence the
separation between the emission and absorption signals in a 21~cm map can
facilitate measurement of the power spectrum due to loops with up to $G\mu \sim 10^{-9}$.

In order to access the detectability, 
we also plot the instrumental noise power spectrum based on the SKA design in Fig.~\ref{fig:llcl_z}.
The instrumental noise power spectrum including the beam effects is given by
\cite{Knox:1995dq}
\begin{equation}
 { N_\ell ^{21}}= 
 { 2 \pi \over t_{\rm obs} \Delta \nu} \left(
{\lambda T_{\rm sys}\over f_{\rm cover} D } 
\right)^2 \exp \left[
 {\frac{\ell(\ell+1)}{\ell_b^2}} \right] ,
\end{equation}
where $T_{\rm sys}$ is the system temperature,
$f_{\rm cover}$ is the a covering fraction of the effective collecting area to the
total collecting area,
$t_{\rm obs}$ is the observation time, $\Delta \nu$ is
the frequency bandwidth, $D$ is the length of the baseline and
$\ell_b$ is given by $\ell _b = 4 \sqrt{\ln 2} /\theta_{fw}$ with
the resolution $\theta _{fw} \sim \lambda /D$.
We set the system temperature 
to the sky temperature in the region
of the minimum emission at high Galactic latitude; $T_{\rm sys} =180 (\nu /180 {\rm
MHz})^{-2.6} \rm K$.
We choose $\Delta \nu$ corresponding to the redshift width $\Delta z =1$
and set $t_{\rm obs} =1~$year.
The current design of
SKA is $ f_{\rm cover} = 0.02$ and $D=
6~$km. However, we take $f_{\rm cover} =0.2$ in this paper.

The noise power spectrum strongly depends on the redshift.
At high redshift ($z=30$), the noise power spectrum totally
overdominates the spectrum due to loops. However, at low redshift
($z=20$), the noise power spectrum in low multipoles becomes smaller than the spectrum for
$G \mu=10^{-8}$,
and is comparable to the one for $G\mu =10^{-9}$.
Therefore, we conclude that
the power spectrum due to loops with $G \mu =10^{-8}$
can be detected via 21~cm observations
with high signal to noise ratio at low
redshifts.

\section{conclusion}

We have investigated the 21~cm signatures induced by cosmic string loops
in this paper. We have taken into account that the loops have the initial relativistic
velocities. A loop with an initial velocity can form a filament
structure.  We have evaluated the gas temperature inside a filament due
to a loop and calculated the differential 21~cm brightness temperature
profile. We have shown that a filament can induce an observable
signal. The brightness temperature can reach $200$~mK for a loop with
$G \mu \sim 10^{-7}$.  

We have also calculated the angular power spectrum of 21~cm fluctuations
due to loops with a scaling loop number density distribution.
The larger $G\mu$ is, the higher the spectrum amplitude due to loops becomes.
The power spectrum is proportional to $\ell$ on large scales, while
it is scale-invariant on small scales.
The scale where the slope of the spectrum changes corresponds to the
typical length scale of the filaments.
The spectrum does not strongly depend on the observation redshift.

We found that the amplitude of the spectrum due to the loops is larger
than the one due to long strings
evaluated in Ref.~\cite{Hernandez:2011ym}.
Therefore, the angular power spectrum of 21~cm fluctuations can give the
limit on $G \mu$ by measuring the spectrum due to loops.
At $z=30$, the angular power spectrum for $G \mu >
10^{-8}$ can dominate the spectrum of the primordial density
fluctuations.  The amplitude due to the primordial fluctuations
becomes smaller as the redshift decreases. Hence, the amplitude of the spectrum
even for $G \mu = 10^{-10}$ can be larger than for the primordial
density fluctuations at $z=20$.  However, first galaxies formed around
$z\sim 20$ produce the larger 21~cm fluctuations through the Ly-$\alpha$
flux and X-ray heating~\cite{2005ApJ...626....1B, Pritchard:2006sq}.

Comparing with the noise power spectrum of the 1 year SKA-like observation,
we have found that the power spectrum for $G \mu = 10^{-8}$ dominates the
noise power spectrum at low multipoles ($\ell <1000$). Therefore, it is
expected that the SKA-like observation can measure the spectrum for $G
\mu > 10^{-8}$ with high signal to noise ratio.

The amplitude of the spectrum depends on the loop number density
distribution which is not completely understood. For example, the
analytical work, Ref.~\cite{Lorenz:2010sm}, suggested scaling loop
distribution with small loop production.  Because of the difference in
the distributions, we found that the angular power spectrum will be
reduced by $\sim 0.1$ with the distribution in
Ref.~\cite{Lorenz:2010sm}.

Since the filaments produced by loops are gravitationally unstable, they
will fragment into beads, which subsequently merge into larger
beads as considered in Ref.~\cite{Shlaer:2012rj}.  Although we have not
taken into account this effect in this paper, the fragmentation of the
filaments decreases the amplitude of the spectrum on large scales and
enhances on small scales.  The evaluation of the power spectrum with the
fragmentation effect requires a detailed study including numerical
simulations.  However, in order to evaluate the fragmentation effect
roughly, we simply assume that the filaments collapse into beads with
the filament width size which corresponds to the fastest growing
instability mode and following mergers of beads makes the length of the
beads ten times larger. Under this assumption, we found that the amplitude of the spectrum is
decreased by 0.1 on large scales and enhanced roughly by 6 on small
scales.

In our calculation of the angular power spectrum, we have only
considered the contributions of small filaments whose virial
temperature is below $10^4~$K.  In massive filaments with the virial
temperature larger than $10^4~$K, stars can be formed and ionize the
surrounding IGM gas~\cite{Shlaer:2012rj}.  The ionized gas can produce
additional large 21~cm fluctuations.  We will study the 21~cm power
spectrum due to massive filaments in future work.


\section*{Acknowledgements}
I thank Eray~Sabancilar and Tanmay~Vachaspati
for their useful comments. 
This work was supported by the DOE. 

\bibstyle{aps}

\end{document}